\documentclass[preprint,showpacs,preprintnumbers,amsmath,amssymb]{revtex4}

\usepackage{graphicx}

\begin{document}


\title{Quantifying Disorder In Point Patterns}
\author{Jeffrey Picka}
 \altaffiliation[Also at ]{University of New Brunswick}
\email{jpicka@unb.edu}
\affiliation{
Department of Mathematics and Statistics, University of New Brunswick, Fredericton NB Canada
}

\pagebreak



\begin{abstract}

Disorder in point patterns can be quantified by means of the complexity, rather than in terms of geometric attributes of pattern structure. A complexity-based disorder-quantifying statistic indicates the practical difficulties associated with modeling processes that produce jammed patterns, particularly with the assessment of model fit and with the simulation of high-intensity hard-core patterns. 

\end{abstract}

\maketitle

\section{Introduction}

Many materials possess a disordered internal structure which cannot be predicted before the material is made. The internal structure of any one specimen of a disordered material of this kind can be modeled as a random sample from an ensemble of disordered structures. The type of disorder present in the material depends both on the structures found in the ensemble and the probability measure defined on the ensemble. If neither of these can be represented by analytical models, then it is necessary to find ways of describing structures which belong to the ensemble and to find ways of distinguishing them from structures which do not belong. In the absence of theory, this requires examination of many specimens of the material and the summarization of common features of the observed disordered structures by means of descriptive statistics. 

Suppose that the structure of a particular specimen of a material can be described by a point pattern arising from a stochastic point process \cite{stoy:1995}. The points could represent the locations of atomic nuclei in an amorphous solid, the centres of monosized spherical grains in a powder, or the centres of spherical inclusions in a composite material. If no two points in a pattern can be closer than some distance $\rho$, then the pattern is a hard-core point pattern. In high-intensity hard-core patterns, monosized spheres attached to the points may be jammed into a rigid arrangement.  Using a descriptive statistic which summarizes aspects of the structure of disordered point patterns, the disorder found in jammed patterns can be shown to be different from the disorder found in general hard-core point patterns. This new statistic is distinct from other statistics which describe how disordered patterns differ from ordered patterns in that it is based on complexity rather than geometry. 

\subsection{Geometric quantification of disorder}

Statistics based on local geometric properties have been used to quantify differences between regular lattice point patterns and disordered point patterns. In some cases a particular lattice is used as the basis of comparison, but in other cases the statistic takes on characteristic values for regular lattice patterns. 

Statistics can be defined based on the Delaunay triangulation of the points \cite{obs}. The edges in the triangulation connect neighbouring points. The orientations of the collection of edges emerging from each point will have distinctive arrangements in regular lattices. To identify these arrangements in arbitrary point patterns, it is necessary to find a way to remove the effects of pattern orientation and of edge labeling. This can be done by evaluating a set of $k^{th}$-order spherical harmonic functions at all edges emerging from a point, and then defining statistics based on averages of these polynomials. Statistics based on $4^{th}$-order spherical harmonics can identify the presence of cubic lattice structure, and statistics based on $6^{th}$-order spherical harmonics can identify the presence of hexagonal  lattice structure. These statistics were originally developed to describe the onset of crystallization in molecular dynamics simulations \cite{ronchetti,steinhardt}, but have been further developed to describe the structure observed in X-ray images of sphere packings \cite{aste} and in simulated sphere packings \cite{tor:2000}.  

Statistics based on triangulation edge orientations at individual points cannot describe near-lattice structures present among small collections of neighbouring points. Statistics based on the shapes of Delaunay simplices can detect these structures \cite{anikeenko:2006}. Statistics of this kind have been been found to be better at tracking crystallization in packings produced by the Jodrey-Tory algorithm \cite{jt:1985,barg:1991} than are statistics based on averages of $6^{th}$-order spherical harmonics \cite{lochmann}. 

Statistics can also be based upon local estimates of intensity \cite{tor:2000} and on the variation among local estimates of intensity \cite{torstil:2003}. If a sphere is defined around an arbitrary point in space, then the variance of the number of other points in that sphere increases proportional to the volume of the sphere (e.g for Poisson point processes) or  increases proportional to the surface area of the sphere (e.g. point lattices, centres of spheres in disordered sphere packings). A statistic based on the rate of increase in variance can be defined and used to rank the disorder of various point arrangements. 

\section{Quantifying disorder by means of complexity}

If a point pattern is a regular lattice, then it is very easy to describe the structure of the pattern, regardless of its size.  A new statistic can be defined by quantifying how difficult it is to communicate the instructions required to reproduce a particular point pattern. It is based on the concept of a minimum description length \cite{rissanen,wallace}.

Let $x$ be a point pattern which may extend throughout all of $\mathbb{R}^d$. Let $W_n$ be a sequence of cubic windows onto the pattern, and let $k_n$ be the number of points in the observed pattern $x_n=x \cap W_n$. 

To give another person enough information to plot the pattern, one could give them the entire list of $d k_n$ point coordinates observed in $W_n$. In the case of a point lattice, much less information is required to reproduce the pattern. Suppose that the pattern must be communicated by means of a general algorithm and an input list of specific information. The general algorithm is intended to reproduce a coordinate list for any subset of a pattern produced by a particular  point process. The algorithm must be formulated on the basis of theory or on the basis of common features observed in many patterns, and cannot be based on attempting to recreate one particular pattern. The list of information is used by the algorithm to create a coordinate list for   the observed pattern. The general algorithm is constructed so that its input list is as short as possible. 
For the pattern $x_n$, the length of its input list will be $L_n$. The disorder-classifying statistic is defined as 
\begin{equation*}
\delta = \lim_{n \rightarrow \infty} L_n/(d k_n).
\end{equation*}
This definition also makes sense for point processes whose patterns are finite.

When there are doubts that a unique limit exists, one can define
\begin{align*}
\delta_{\ell} &= \liminf_{n \rightarrow \infty} L_n/(d k_n) \\
\delta_{u} &= \limsup_{n \rightarrow \infty} L_n/(d k_n)
\end{align*}

The limit $\delta$ is defined for particular patterns rather than for the entire process. To define a measure of disorder for a point process, the ensemble average of $\delta$ over all patterns consistent with that process would be used.  It would not make sense to use the infimum of $\delta$ over all realizations of a process, since point lattices are among the possible realizations for some highly disordered point processes. 


\subsubsection{Example: A cubic array}

Consider a cubic lattice packing of monosized spheres in $\mathbb{R}^3$. To reconstruct the point lattice generated by the centres of these spheres, it is necessary to know the location of one sphere centre, the locations of two of its neighbours whose centres do not lie on a straight line including the centre of the initial sphere, the location of one vertex of $W_n$, two direction vectors to orient $W_n$, and the length of one of the sides of $W_n$. A general algorithm can be written which takes this list of numbers and finds all sphere centres lying within $W_n$. As $n$ increases, all that needs to change is the size of the side length of $W_n$ and possibly its location vertex. The list contains the same number of items for all $W_n$, and so $\delta=0$. If an algorithm could be found requiring a shorter input list, this argument still shows that $\delta$ takes on its smallest possible value. 

\subsubsection{Example: A Markovian hard-core point process}

The simplest and most analytically tractable form of hard-core point process model is the hard-core Strauss process \cite{strauss}. It is a Markovian point process, in that the location of any one point in a realization depends only pairwise on the positions of its immediate neighbours. This conditions renders the hard core Strauss process analytically tractable, but also puts restrictions on the patterns which lie in the ensemble from which the process samples. 

If $x_n$ is a realization of a hard core Strauss process, then the only way to reconstruct the pattern  within $W_n$ is to print out the original list of point locations. No further reduction in list size is possible, since no one point location is completely determined by any other point location. This gives a constant value of $\delta=1$ for all realizations. 

Since realizations of a Markovian hard-core process would be produced by a Gibbs process simulation algorithm \cite{molwag}, the full list of locations could be generated by defining the general algorithm to be the simulation algorithm for the process and its random number generator. The realization could then be produced by a list that gives the seed of the generator and the size, location, and orientation of $W_n$. Then, $\delta=0$ as in the case of the point lattice. This type of argument is forbidden, since the simulation program and the seed cannot be inferred by looking at a sample of patterns or by analysis of the un-normalized density of the process.

\subsubsection{Example: A sphere packing} 

Consider a pattern defined by the centres of spheres in a monosized sphere packing in which every sphere is jammed rigidly into place by its neighbours. Within $x_n$, remove a point which is at the centre of a sphere that lies entirely within $W_n$. Use an algorithm which lists every other sphere centre, and then searches for a region in space where the missing sphere could be made to fit. The algorithm then generates the location of the centre of that missing sphere. By omitting the centre of the removed sphere from the list, the list is now $d$ entries shorter in length.  

Given a large enough packing, a replacement algorithm could rebuild the packing from a shorter list. Consider an algorithm that  identifies locations where isolated spheres are missing from a packing, and then supplies the coordinates of the missing sphere centres. This algorithm could be supplied with a list of centre locations which omits a subset of points, chosen so that each removed sphere sits entirely within $W_n$ and has no neighbours that are also neighbours of another removed sphere. These removed sphere centres would form a  homogeneous subset of the original pattern, and their number would increase proportionally to the size of $W_n$. Thus,  $\delta_u < 1$. This algorithm would not produce the shortest list possible, but it would be very difficult to determine how small the shortest list could be. It would be necessary to find an algorithm that could replace many missing spheres, and decide if a list could be used to reconstruct two different patterns. It would then be necessary to examine all $2^k$ subsets of the $k$ spheres in the realization and then to check each subset for reconstructability. 

When physical packings of spherical objects are produced by pouring spheres from one container into another, the patterns that result are random because the physical process that generates the patterns is sensitive to uncontrollable initial and boundary conditions. Given any sufficiently large bounded container, a pouring process can fill it with  many different packed patterns. As a result, if $W_n$ is sufficiently large and bisected by a plane parallel  to a pair of its sides, then removal of all spheres in one half of $W_n$ will not allow a unique reconstruction of the missing part of the pattern. Thus $\delta_{\ell}>0$ and, if $\delta$ is well-defined, then $0 < \delta < 1$. 

\section{Discussion} 

Different types of disorder can be classified by geometry-based statistics which measure differences from order. Some statistics based on spherical harmonics can track the changes in ensemble structure over time for a packing simulator \cite{lochmann}, but others are unable to clearly distinguish between realizations of different physical packing processes \cite{aste}. The variance-based statistic may be useful when the sampled volume is very large, but it assigns different values to the cubic and hexagonal close-packed lattices \cite{torstil:2003}. The geometry-based statistics describe disordered patterns in a different way from more conventional descriptive statistics such as pair correlation functions, and so they can be used to enlarge the set of statistics which can be used to describe differences between the patterns in different ensembles. 

The complexity-based statistic describes both conceptual and practical problems associated with the modeling of disorder, particularly in the case of modeling high-intensity hard-core point processes. For a packed pattern $\delta_{\ell}>0$, since the number of distinct disordered jammed patterns increases with volume. For large systems, any regular lattice structure is just one among many possible jammed patterns. The rigidity of the jammed patterns forces $\delta_u<1$. The rigidity implies that the jammed patterns form a collection of many isolated patterns within the much larger collection of hard-core patterns. 

The distinction between the ensemble of jammed patterns and the much larger ensemble of hard-core patterns has important implications for modeling. It is possible to formally define a stochastic point process over an ensemble containing some of the hard-core point processes, provided that a Markovian assumption is made to restrict point interactions. This assumption takes the form of a restriction that the partition function contain no interaction terms other than those representing pairwise interactions \cite{ruelle}. This assumption is required in order to formally state the partition function, and is also required in order to make any formal analysis of the model without resorting to examining collections of patterns generated by the process. There are no such formally-defined point process models on the ensemble of packed patterns. This makes formal analysis and modeling of these patterns impossible, and forces any analysis of the model to be based on a sample of patterns from the process. 

If jammed arrangements of spheres are to be modeled, then there must be some form of probability measure defined on an ensemble of jammed patterns. Since different physical preparation methods result in sampling from different subsets of the ensemble of all jammed patterns  \cite{scott:1960}, it cannot be assumed that an arbitrary packing model will sample from the right ensemble. At present, all models for packings are simulation routines which assemble jammed patterns by some relatively simple algorithm \cite{jt:1985,barg:1991,lubstil:1990,lubstil:1991}. None of these programs are based on faithful modeling of the physics of packing formation. While the DEM models \cite{zhu,zhu:2008} appear at first glance to be physically faithful, they are based on unverifiable assumptions about grain interactions and, if they are deterministic, they can at best provide a sample of size 1 from the ensemble of packed patterns. Without a line of experimentally supported physical reasoning which can justify that a particular simulation routine represents the physics of a packing formation process, it is necessary to assess the fit of any proposed model by comparing patterns from the model to patterns arising from replications of the physical process. 


The distinction between ensembles for packing processes and for hard-core point processes also has implications for the general simulation of hard-core point processes. If hard spheres are attached to every point in a hard-core pattern, then there is an upper limit of volume fraction that can be obtained by birth-death simulation algorithms for hard-core Markov point processes \cite{geyer:1999}. If randomly-placed but not jammed non-overlapping spheres model a physical system and the volume fraction of spheres exceed this limit, then hard-core Markov point processes cannot be used as the basis of simulating these materials or as the basis for analyzing their properties. Instead, these materials need to be simulated from jammed patterns in which every point is randomly perturbed. These algorithms must run through many re-arrangement cycles to be certain that the initial jammed pattern does not dominate the final simulated pattern.  


\end{document}